\documentstyle[12pt]{article}                                      
                                      
\newlength{\dinwidth}                                      
\newlength{\dinmargin}                                      
\def\lapproxeq{\lower .7ex\hbox{$\;\stackrel{\textstyle                                      
<}{\sim}\;$}}                                      
\def\gapproxeq{\lower .7ex\hbox{$\;\stackrel{\textstyle                                      
>}{\sim}\;$}}                                      
\def\be{\begin{equation}}                                      
\def\ee{\end{equation}}                                      
\def\bea{\begin{eqnarray}}                                      
\def\eea{\end{eqnarray}}

\begin{document}                                      
\titlepage                                      
\begin{flushright}                                      
DTP/99/32 \\                                      
March 1999 \\                                      
\end{flushright}                                      
                                      
\vspace*{2cm}                                      
                                      
\begin{center}                                      
{\Large \bf Predictions for dijet production in DIS} \\       
       
\vspace*{0.7cm}       
{\Large \bf using small $x$ dynamics}                                      
                                      
\vspace*{1cm}                                      
J.~Kwiecinski$^{a,b}$, A.D.~Martin$^b$ and A.M.~Stasto$^{a,b}$ \\                                      
                                     
\vspace*{0.5cm}                                      
$^a$ H.~Niewodniczanski Institute of Nuclear Physics, ul.~Radzikowskiego 152,                          
Krakow, Poland \\                         
$^b$ Department of Physics, University of Durham, Durham, DH1 3LE \\                                    
\end{center}                                      
                                      
\vspace*{2cm}                                      
                                      
\begin{abstract}                                      
We study the properties of dijet production in deep inelastic scattering using a unified       
BFKL/DGLAP framework, which includes important subleading $\ln (1/x)$       
contributions.  We calculate the azimuthal decorrelation between the jets.  We       
compute the cross section for dijet production as a function of $Q^2$ and the jet       
transverse momentum, as well as calculate the total dijet rate.  We compare the       
predictions with HERA data.      
\end{abstract}                                     
                             
\newpage                                     
Dijet production in high energy deep inelastic electron-proton scattering can expose       
properties of small $x$ behaviour in QCD, as can be seen from Fig.~1.  In the       
dominant $\gamma^* g \rightarrow q\bar{q} \rightarrow$~dijet subprocess, the       
incoming gluon can have sizeable transverse momentum accumulated from diffusion       
in $k_T$ along the gluon chain \cite{AGKM,FR}.  The value of the transverse    
momentum, and hence the azimuthal decorrelation between the jets, increases with    
decreasing $x$.  That is the jets are no longer back-to-back since they must balance    
the appreciable transverse momentum $k_T$ of the incoming virtual gluon.  The    
azimuthal decorrelation from the back-to-back configuration $\phi = \pi$ is therefore a    
measure of $k_T$ and may be expected to be an indicator of the diffusion along the    
BFKL chain.  Clearly to obtain a reliable measure we must avoid the infrared region    
$k_T \simeq 0$ (that is $\phi \simeq \pi)$.  However, as will be seen, we are able to   
make an essentially parameter-free prediction of the integrated dijet production rate (at   
the    
parton level).      
      
The description of dijet production is based on the unfolded $k_T$ factorization       
formula for structure functions \cite{KTF1,KTF2}, which exposes the unintegrated    
gluon distribution $f       
(x_g, k_T^2)$ more locally than the structure functions themselves.  The calculation       
goes beyond the conventional (fixed order) QCD-improved parton model, which is       
known to underestimate the dijet rate \cite{P}.  We will find that the approach       
driven by the unintegrated gluon distribution gives an enhancement in the predicted       
rate and hence an improvement in comparison with the data.      
      
The differential cross section for producing two jets of transverse momenta $p_{1T}$       
and $p_{2T}$ in deep inelastic scattering is      
\be      
\label{eq:a1}      
\frac{d \sigma}{dx dQ^2 d \phi dp_{1T}^2 dp_{2T}^2} \; = \; \frac{4 \pi \alpha^2}{x       
Q^2} \left [ \left (1 - y - \frac{y^2}{2} \right ) \: \frac{d F_T}{d \phi dp_{1T}^2       
dp_{2T}^2}  \: + \: (1 - y) \: \frac{d F_L}{d \phi dp_{1T}^2 dp_{2T}^2} \right ]      
\ee      
where, as usual, the deep inelastic variables $Q^2 = -q^2, x = Q^2/2p \cdot q$ and $y       
= p \cdot q/p \cdot p_e$ where $p_e, p$ and $q$ are the four momenta of the incident       
electron, proton and virtual photon respectively, see Fig.~1.  The differential structure       
functions are obtained from the $k_T$ factorization prescription by unfolding the       
integrations over $p_{1T}^2, p_{2T}^2$ and the azimuthal angle $\phi$ between the       
$q$ and $\bar{q}$ jets.      
      
It is convenient to use a Sudakov decomposition of the jet four momenta      
\bea      
\label{eq:a2}      
p_1 & = & (1 - \beta) q^\prime \: + \: \alpha_1 p \: + \: \mbox{\boldmath $p$}_{1T}       
\nonumber \\      
& & \\      
p_2 & = & \beta q^\prime \: + \: \alpha_2 p \: + \: \mbox{\boldmath $p$}_{2T}       
\nonumber      
\eea      
where $q^\prime = q + xp$ and $p$ are the basic lightlike momenta.  Since the jets       
are on-mass-shell      
\be      
\label{eq:a3}      
\alpha_1 \; = \; \left ( \frac{p_{1T}^2 + m_q^2}{(1 - \beta) Q^2} \right ) x, \quad\quad       
\alpha_2 \; = \; \left ( \frac{p_{2T}^2 + m_q^2}{\beta Q^2} \right ) x      
\ee      
where $m_q$ is the mass of the quark.  The $k_T$ factorization formula for the       
differential structure functions is      
\be      
\label{eq:a4}      
\frac{dF_i}{d \phi dp_{1T}^2 dp_{2T}^2} \; = \; \sum_q \: \int_0^1 \: d \beta \: {\cal       
F}_i^q (\beta, p_{1T}^2, p_{2T}^2, \phi, Q^2) \: \frac{f (x_g, k_T^2)}{k_T^4}      
\ee      
with $i = T, L$.  The function $f (x_g, k_T^2)$ is the unintegrated gluon distribution    
describing the gluon chain in Fig.~1.  The variables $x_g$ and $k_T$ are the    
longitudinal momentum fraction and the transverse momentum, relative to the proton,   
carried by the gluon which couples to the $q\bar{q}$ jet pair.  They are given by   
\bea      
\label{eq:a5}      
x_g & = & x \: + \: \alpha_1 \: + \: \alpha_2 \; \quad\quad (\gapproxeq [1 + 4     
p_{iT}^2/Q^2] x)       
\nonumber \\      
& & \\      
k_T^2 & = & p_{1T}^2 \: + \: p_{2T}^2 \: + \: 2p_{1T} p_{2T} \cos \phi, \nonumber      
\eea      
see Fig.~1.  The functions ${\cal F}_i$, which describe the virtual photon-gluon       
fusion subprocess $\gamma g \rightarrow q\bar{q}$, are      
\bea      
\label{eq:a6}      
{\cal F}_T^q & = & e_q^2 \alpha_S (k_T^2) \frac{Q^2}{8 \pi^2} \left \{ \biggl       
[\beta^2 + (1 - \beta)^2 \biggr ] \left ( \frac{p_{1T}^2}{D_1^2} +       
\frac{p_{2T}^2}{D_2^2} + \frac{2 p_{1T} p_{2T} \cos \phi}{D_1 D_2} \right ) +       
m_q^2 \left ( \frac{1}{D_1} - \frac{1}{D_2} \right )^2 \right \} \\      
& & \nonumber \\      
\label{eq:a7}      
{\cal F}_L^q & = & e_q^2 \: \alpha_S (k_T^2) \: \frac{Q^4}{2 \pi^2} \: \beta^2 (1 -       
\beta)^2 \: \left ( \frac{1}{D_1} \: - \: \frac{1}{D_2} \right )^2      
\eea      
where $e_q$ is the charge of the quark and where the denominators      
\be      
\label{eq:a8}      
D_i \; = \; p_{iT}^2 \: + \: m_q^2 \: + \: \beta (1 - \beta) Q^2.      
\ee      
The unintegrated gluon distribution $f (x_g, k_T^2)$ is obtained by using a unified       
BFKL/DGLAP formalism, which includes sub-leading $\ln (1/x_g)$    
contributions\footnote{Next-to-leading order (NLO) $\ln (1/x)$ corrections are known    
to be large \cite{NLO1,NLO2}, and imply the need for an all-order resummation.  We    
are able to include a major part of the all-order resummation by imposing a    
consistency constraint on the BFKL kernel \cite{CC}.  The constraint requires the    
virtuality of the exchanged gluons along the chain to be dominated by their transverse    
momentum squared.  Although the constraint contributes at all orders, its NLO    
contribution gives about 70\% of the exact NLO result for the BFKL Pomeron    
intercept.}, to       
fit to the inclusive $F_2$ structure function data \cite{STASTO}.  In this way $f (x_g,       
k_T^2)$ is determined for $k_T^2 > k_0^2$ where $k_0^2 = 1$~GeV$^2$.  Formally       
the integration limits for $\beta$ in (\ref{eq:a4}) are 0 and 1, but they are constrained       
by the condition $x_g < 1$.  Moreover the lower limit on $k_T$ in the determination       
of the unintegrated gluon means that we cannot predict the azimuthal decorrelation       
between the jets in the near back-to-back domain $\phi \simeq \pi$.  Our decorrelation       
predictions are limited to the region      
\be      
\label{eq:a9}      
1 \: + \: \cos \phi \: > \: k_0^2/2 p_0^2      
\ee      
where $p_0$ is the minimal value of the transverse momentum of an outgoing jet in       
the dijet system.      
      
Predictions for the azimuthal decorrelation are shown in Fig.~2.  They are compared       
with the measurements made using the ZEUS detector which so far have been       
presented in the thesis of Przybycien \cite{P}.  We use the parton level data obtained       
with the $k_T$ jet-finding algorithm.  The predictions use the same cuts as the data;       
that is $Q^2 > 8$~GeV$^2$, outgoing electron energy    
$< 10$~GeV, $p_T$(jet) $< 4$~GeV, $-2 < \eta ({\rm lab}) < 2.2$ and $\eta ({\rm    
HCM}) > 0$, where the       
pseudorapidities $\eta$ refer to the HERA and hadron centre-of-mass frames      
respectively.  In       
each $x$ bin the data for the $\phi$ distribution are normalized to unity, and the      
predictions are normalized to the       
data point centred at $\phi = 2.55$ radians $(\phi = 146^\circ$).  We see that there is       
satisfactory agreement between the predictions and the shape of the observed $\phi$      
distributions.  It should, however, be       
noted that, due to the cuts, eq.~(\ref{eq:a5}) implies that $x_g \gapproxeq 9 x$.  Thus       
the observed $x$ bins do not fully expose the small $x_g$ domain.  Hence the       
broadening of the azimuthal decorrelation, although visible in the predictions, is       
quantitatively marginal.  The $x_g$ distribution of the data is compared with the      
predictions in Fig.~3(a).  We see these data sample the gluon in the region $x_g      
\simeq 10^{-2}$.   Not surprisingly, in this $x_g$ domain the $\phi$ distribution does    
not give a definitive test of the underlying dynamics.  Indeed Monte Carlos, which do   
not embody BFKL effects, can also describe the $\phi$ distribution reasonably well    
\cite{P}.   
      
We emphasize that the calculation of dijet production is essentially parameter free.    
Moreover it applies to the full kinematic domain since it is based on an unintegrated    
gluon which is obtained from a unified BFKL/DGLAP approach.  The parameters    
which enter the determination of the unintegrated gluon are       
completely specified by the fit to the $F_2$ structure function data \cite{STASTO}.        
Thus we can make an absolute comparison with the measured cross section for dijet       
production.  Fig.~3(b) shows the comparison as a function of $Q^2$.  At large $Q^2$       
there is excellent agreement.  However as $Q^2$ decreases the prediction, with its       
weaker $Q^2$ dependence, falls below the data.  The reason is that for $Q^2 \ll       
4p_{iT}^2$ the denominators $D_i$ of (\ref{eq:a8}) are dominated by $p_{iT}^2$       
and hence the calculated cross section depends only weakly on $Q^2$.  There is a       
natural explanation of the discrepancy in Fig.~3(b).  Dijets may also be produced in       
the photon hemisphere from the higher order contribution in which one of the jets is a       
gluon emitted from the quark box, that is $\gamma g \rightarrow gq (\bar{q})$ or       
$g\bar{q} (q)$ with a spectator $\bar{q}$ or $q$ of small $p_T$.  We expect a more       
rapid $Q^2$ fall-off from such a contribution.      
      
In order to calculate the $x_g$ and $Q^2$ distributions of Fig.~3 we integrate over       
the entire $k_T^2$ range of the gluon.  The infrared contribution from $k_T^2 <       
k_0^2$ is estimated using the prescription of ref.~\cite{STASTO}.  That is we use the       
strong-ordering approximation $k_T^2 \ll p_T^2$ and express the corresponding       
integrals in terms of the integrated gluon distribution $g (x_g, k_0^2)$ at scale       
$k_0^2$.  This non-perturbative input is determined by the fit to the $F_2$ data in a       
self-consistent way in the unified BFKL/DGLAP framework \cite{STASTO}.      
      
In Table 1 we make the comparison of the data and the predictions of the dijet cross       
section for different values of the minimal $p_T$ of the jets.  The calculation       
reproduces 70--80\% of the observed rate.   To put this comparison in context, we note    
that the Mepjet Monte Carlo predicts a dijet cross section for $p_T$(jet) $> 4$~GeV    
of 2.8\footnote{This number corresponds to Mepjet 2.0 with $Q$ as the scale.  If    
$p_T$ is taken to be the scale the cross section drops to 2.4~nb \cite{P}.} or 2.6~nb    
according to whether GRV \cite{GRV} or MRSA \cite{MRSA} partons were used    
\cite{P}.  The latter set of partons have an integrated gluon more compatible with that    
used for our analysis.  Thus the inclusion of small $x$ contributions are seen to    
enhance the cross section, although the prediction of 3.2~nb is still below the    
measured value of 3.9~nb.   
   
\begin{table}[htb]   
\begin{center}   
\begin{tabular}{|c|c|c|} \hline   
$p_T$ & $\sigma$ (expt) & $\sigma$ (theory) \\    
in GeV & in nb & in nb \\ \hline   
4 & 3.9 & 3.2 \\   
5 & 2.6 & 1.8 \\   
6 & 1.6 & 1.1 \\   
7 & 1.0 & 0.7 \\   
8 & 0.6 & 0.5 \\ \hline   
\end{tabular}   
\end{center}   
\caption{The comparison of the measured \protect\cite{P} and theoretical integrated    
dijet production cross sections at the parton level for different $p_T$ cuts on the jet    
transverse momenta in the hadronic centre-of-mass frame.}   
\end{table}   
   
In principle dijet production appears to offer an opportunity to study the $k_T$    
diffusion property of the BFKL gluon $f (x_g, k_T^2)$.  In practice the cuts,    
necessary on the transverse momentum of the jets, curtail the small $x_g$ \lq\lq    
reach\rq\rq~of HERA, see (\ref{eq:a5}).  Fortunately our unified BFKL/DGLAP    
approach is not restricted to small $x_g$, and gives a satisfactory description of the    
azimuthal decorrelations of the jets.  Not surprisingly, the description is not unique    
and several standard Monte Carlos are known to also be able to accommodate the    
decorrelation data.  The $\phi$ distribution, in the presently accessible kinematic    
domain, cannot therefore be regarded as a discriminator of the underlying small $x$    
dynamics.  However our BFKL/DGLAP framework (with subleading $\ln (1/x)$    
contributions) does give an enhancement of the dijet rate, although the prediction still    
falls short of the observed cross section.  Moreover by comparing the predicted $Q^2$    
dependence of the cross section with the data we are able to reveal the potential source    
of the remaining discrepancy. \\   
   
\noindent {\large \bf Acknowledgements}   
   
We thank Brian Foster and Maciej Przybycien for discussions about the data.  JK and    
AMS thank the Physics Department and Grey College of the University of Durham for    
their warm hospitality.  AMS thanks Foundation for Polish Science for support.
This work was supported in part by the UK Particle Physics    
and Astronomy Research Council, by the Polish State Committee for Scientific    
Research (grant no.~2~P03B~089~13) and by the EU Fourth Framework Programme    
TMR Network \lq QCD and Particle Structure\rq~(contract FMRX-CT98-0194,    
DG12-MIHT).

\newpage   
\noindent {\large \bf Figure Captions}   
\begin{itemize}   
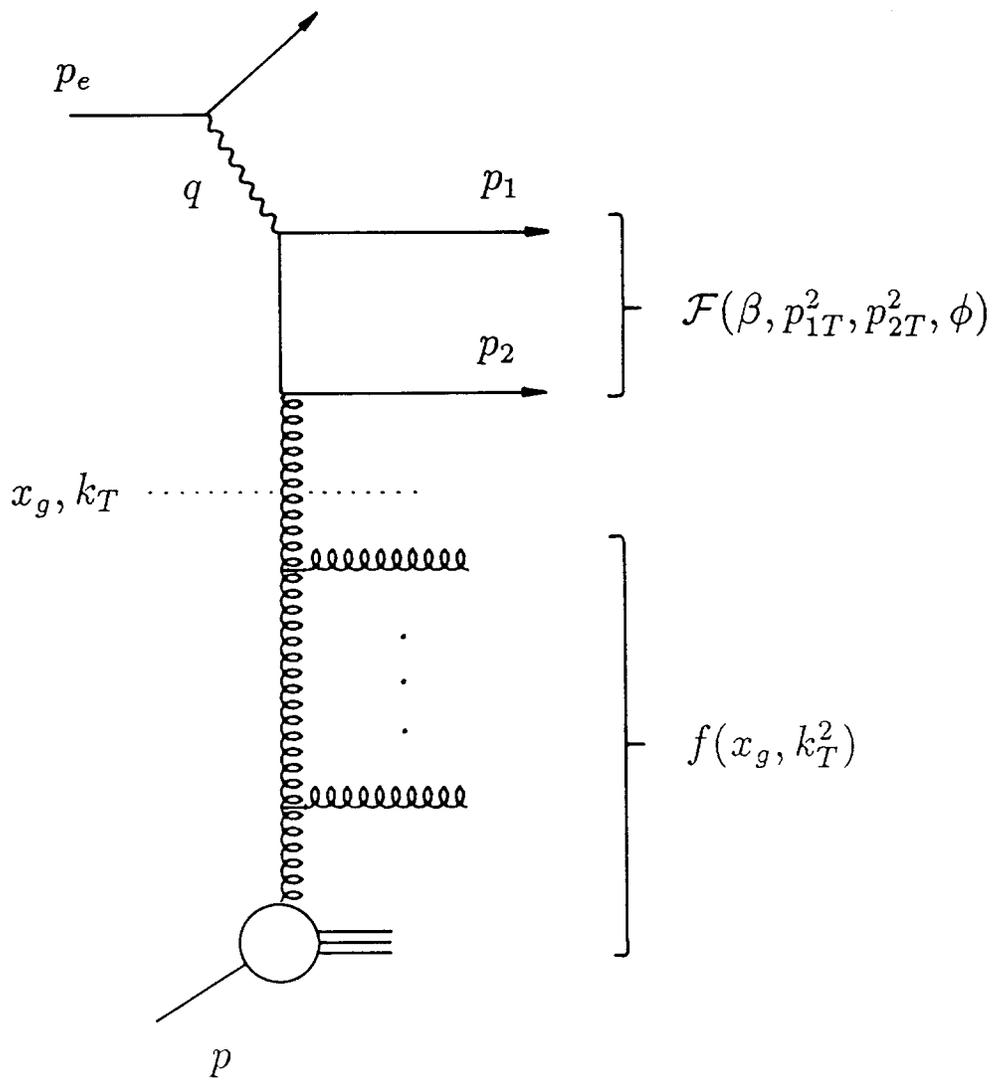
\item[Fig.~1] Diagrammatic representation of dijet production in deep inelastic    
scattering via the photon-gluon subprocess.  It defines the kinematics and dynamical    
quantities entering the $k_T$ factorization formula (\ref{eq:a4}).   
   
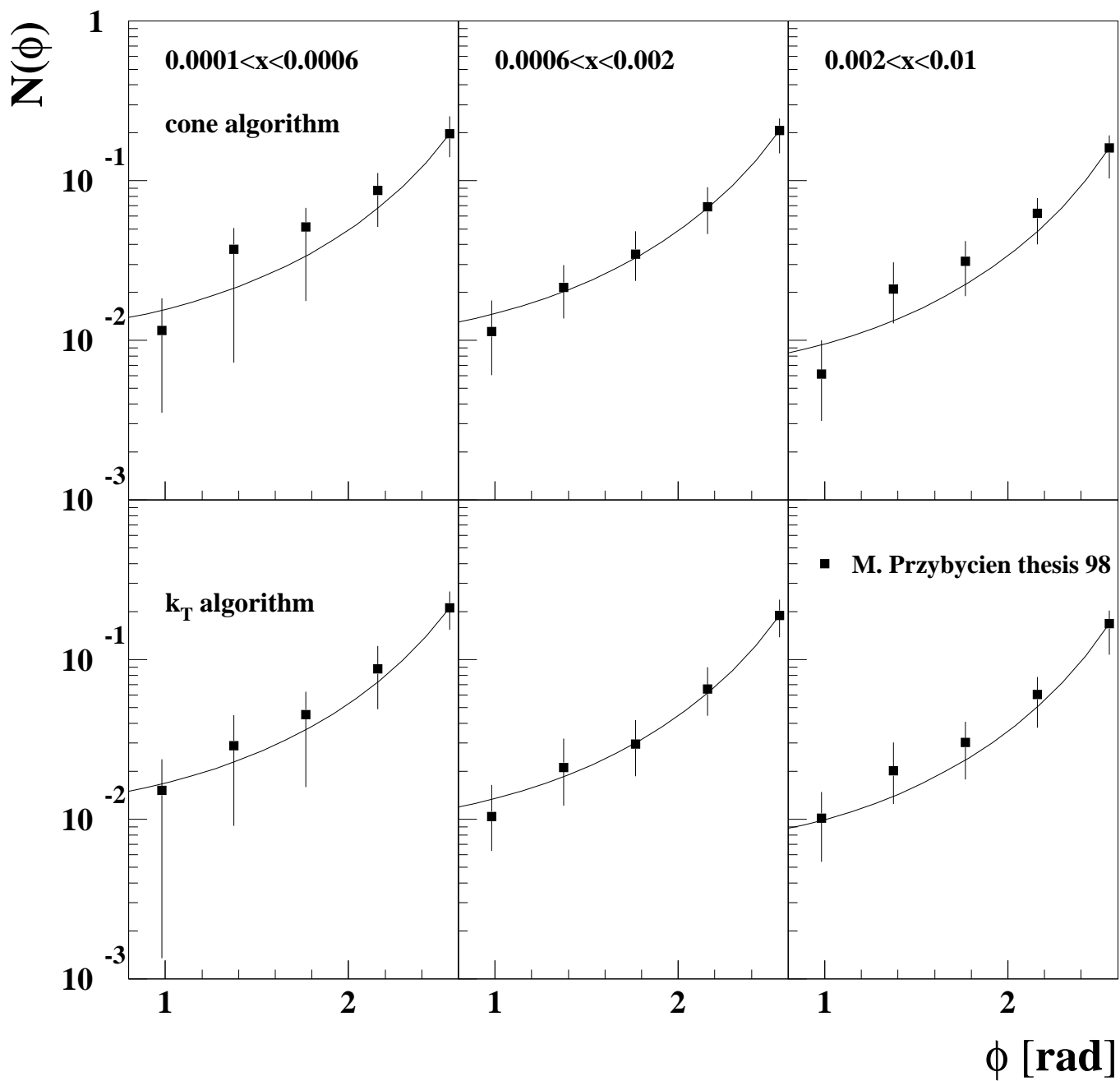
\item[Fig.~2] Theoretical predictions for the distribution with respect to the azimuthal    
angle $\phi$ between the $q$ and $\bar{q}$ jets, compared to the data of    
ref.~\cite{P}, for three different intervals of Bjorken $x$.  The data are at the parton    
level and were obtained using the $k_T$ jet-finding algorithm.  The predictions are    
normalised to the data point at $\phi = 2.55$ radians.   
   
\item[Fig.~3] Theoretical predictions for (a) the $x_g$ dependence, and (b) the $Q^2$    
dependence, of the dijet production cross section compared to the parton-level data of    
ref.~\cite{P}.   
   
\end{itemize}   
                       

\newpage      
\begin{thebibliography}{xx}      
\bibitem{AGKM} A.J.~Askew, D.~Graudenz, J.~Kwiecinski and A.D.~Martin,    
Phys.~Lett.~{\bf B338} (1994) 92.   
%
\bibitem{FR} J.R.~Forshaw and R.G.~Roberts, Phys.~Lett.~{\bf B335} (1994) 494.   
%
\bibitem{KTF1} S.~Catani, M.~Ciafaloni and F.~Hautmann, Phys.~Lett.~{\bf B242}    
(1990) 97; Nucl.~Phys.~{\bf B366} (1991) 135; S.~Catani and F.~Hautmann,    
Nucl.~Phys.~{\bf B427} (1994) 475; M.~Ciafaloni and D.~Colferai, {\tt    
hep-ph/9806350}.   
%
\bibitem{KTF2} J.C.~Collins and R.K.~Ellis, Nucl.~Phys.~{\bf B360} (1991) 3.   
%
\bibitem{P} M.~Przybycien, Ph.D.~thesis \lq\lq Two jets production in neutral current   
deep inelastic $e^+ p$ interactions at 300~GeV c.m.s.~energy\rq\rq, Krakow,   
Dec.~1998.      
%
\bibitem{NLO1} V.S.~Fadin, M.I.~Kotskii and R.~Fiore, Phys.~Lett.~{\bf B359}    
(1995) 181; V.S.~Fadin, M.I.~Kotskii and L.N.~Lipatov, {\tt hep-ph/9704267};    
V.S.~Fadin, R.~Fiore, A.~Flachi and M.I.~Kotskii, Phys.~Lett.~{\bf B422} (1998)    
287; V.S.~Fadin and L.N.~Lipatov, {\tt hep-ph/9802290}; Phys.~Lett.~{\bf B429}    
(1998) 127; V.S.~Fadin, {\tt hep-ph/9807527}, {\tt hep-ph/9807528}; M.~Ciafaloni    
and G.~Camici, Phys.~Lett.~{\bf B386} (1996) 341; {\bf B412} (1997) 396; {\bf    
B417} (1998) 390 (E); Phys.~Lett.~{\bf B430} (1998) 349; M.~Ciafaloni, {\tt    
hep-ph/9709390}.   
%
\bibitem{NLO2} D.A.~Ross, Phys.~Lett.~{\bf B431} (1998) 161; \\   
G.P.~Salam, JHEP~{\bf 9807} (1998) 019.   
%
\bibitem{CC} J.~Kwiecinski, A.D.~Martin and P.J.~Sutton, Z.~Phys.~{\bf C71}    
(1996) 585.   
%
\bibitem{STASTO} J.~Kwiecinski, A.D.~Martin and A.M.~Stasto, Phys.~Rev.~{\bf       
D56} (1997) 3991.     
%
\bibitem{GRV} M.~Gl\"{u}ck, E.~Reya and A.~Vogt, Z.~Phys.~{\bf C67} (1995)    
433.   
%
\bibitem{MRSA} A.D.~Martin, R.G.~Roberts and W.J.~Stirling, Phys.~Rev.~{\bf    
D50} (1994) 6734.   
\end{thebibliography}
\end{document}